\documentstyle[12pt,aasms4]{article}

\begin{document}

\title{ PKS 0405--385: the smallest radio quasar?}
\author{L.~Kedziora-Chudczer\altaffilmark{1,2}, D.~L.~Jauncey\altaffilmark{2}, M.~H.~Wieringa\altaffilmark{2}, M.~A.~Walker\altaffilmark{1}, G.~D.~Nicolson\altaffilmark{3}, J.~E.~Reynolds\altaffilmark{2}, A.~K.~Tzioumis\altaffilmark{2}}
\affil{1. Special Research Centre for Theoretical Astrophysics, School of Physics A28, Sydney University, NSW 2006, Australia}
\affil{2. Australia Telescope National Facility, CSIRO, PO Box 76, Epping, NSW 2121, Australia}
\affil{3. Hartebeesthoek Radio Astronomy Observatory,
P.O.~Box~443, Krugersdorp 1740, Transvaal, South Africa}
\authoremail{lkedzior@atnf.csiro.au}

\begin{abstract}
We have observed profound variability in the radio
flux density of the quasar PKS~0405--385 on timescales of less than an
hour; this is unprecedented amongst extragalactic sources.
If intrinsic to the source, these variations would imply a
brightness temperature ${\rm T_B\sim10^{21}\;K}$, some nine
orders of magnitude larger than the inverse Compton limit
for a static synchrotron source, and still a million times
greater than can be accommodated with bulk relativistic
motion at a Lorentz factor $\gamma\sim10$. The variability
is intermittent with episodes lasting a few weeks to months. 

Our data can be explained most sensibly as interstellar
scintillation of a source component which is $<5$~$\mu$arcsec
in size --- a source size which implies a brightness temperature
${\rm T_B>5\times10^{14}\;K}$, still far above the inverse
Compton limit. Simply interpreted as a steady, relativistically
beamed synchrotron source, this would imply a bulk Lorentz factor
$\gamma\sim10^3$. 
\end{abstract}

\keywords{galaxies: variability and quasars --- quasars: individual (PKS 0405-385), radiation mechanisms and gravitational lensing --- radio continnum: quasars --- ISM: scattering}

\section{Introduction}
Most quasars display properties which are entirely
consistent with their radio flux density being caused by synchrotron
emission (e.g. \markcite{Begelman84} Begelman et al. 1984). For high
source intensities a second radiation process -- inverse-Compton
scattering -- must also be an important energy-loss mechanism
for the relativistic electrons. This
competing process limits the synchrotron intensity of any static radio source to
brightness temperatures ${\rm T_B<10^{12}\;K}$ \markcite{Kellermann}(Kellermann \& Pauliny-Toth 1969). 
A brightness temperature can be deduced from flux density variations,
the time scale of which is taken to indicate the size 
of the source. Thus, much interest has been aroused in recent years by
the phenomenon of intraday variability (IDV) of radio sources,
where the inferred brightness temperatures are many orders
of magnitude higher than the inverse Compton limit --- see \markcite{Wagner} Wagner \& Witzel (1995) for a review. It is not yet understood how such
high apparent brightness temperatures arise.

In order to investigate the IDV phenomenon further,
we carried out a survey of 125 compact southern
radio sources at four frequencies (1.4, 2.4, 4.8 and 8.6~GHz), using
the Australia Telescope Compact Array (ATCA). Here we present our
observations of one source, PKS~0405--385, which stood out
from all others by virtue of the extreme rapidity and large
amplitude of its variations (\markcite{Kedziora} Kedziora-Chudczer et al. 1996).
Our data are described in \S2, and the theoretical
deductions presented in \S3.
\section{Observations}

PKS 0405--385 is a compact, inverted-spectrum radio quasar
with a redshift z=1.285 (\markcite{Peterson} Peterson \& Bolton 1972; \markcite {Veron}V\'eron et~al. 1990). 
Its optical spectrum
exhibits strong, broad emission lines, with an absorption
system at z=0.875 (\markcite{Hunstead} Hunstead 1996). Available VLBI
data indicate that the source has no radio structure on scales larger than
1~milli-arcsec. Its linearly polarized flux density does not exceed 3\%
of the total intensity for any of our observed frequencies.  

Intraday variability of PKS 0405--385 was first observed in
1993 November during our ATCA IDV survey. At that time
the strongest and fastest flux density changes were observed at 8.6 GHz
 (17\% peak-to-peak
within 24 hours). In 1994 May we found 20\% peak-to-peak flux density excursions in 2 hours (our sampling
interval) at
2.4 GHz. We monitored PKS 0405--385 continuously with the ATCA on
1996 June 7 and realized immediately the unique nature of its 
flux density
variations. Figure 1 shows the flux density variations of PKS~0405--385, as
measured by the ATCA at four frequencies over 90 hours. Data were collected
simultaneously for two frequency pairs: either 1.4 / 2.4~GHz or 4.8 / 8.6~GHz. 
We switched between these two pairs every 10 min. 

Independent confirmation of the rapid variability of
PKS 0405--385 came almost immediately from Hartebeesthoek Radio Astronomy Observatory
(HARTRAO) measurements at 5.0 and 8.4~GHz. These data are shown in Figure~2.
Within a week, further confirmation of our
result was provided by observations with the NRAO's
Very Large Array (VLA)\footnote{The National Radio Astronomy Observatory
is operated by Associated Universities Inc., under cooperative agreement
with the National Science Foundation of the USA}, at 8 and 22~GHz.

At 4.8 and 8.6 GHz the variations are quasi-sinusoidal, and there is strong correlation
between these two frequencies, with no detectable time lag. The time scale of the variations is much greater at
2.4 and 1.4 GHz, and the amplitude of variability is smaller.
The fractional rate of change of flux density for PKS~0405--385 is over 50\% per hour at
4.8~GHz. For comparison the swiftest
observed flux density change in the IDV source 0917+624 at 2.7 GHz (\markcite{Qian} Qian et al. 1991) was 1.5\% per hour while the fastest variation
seen in OJ 287 in 11 days of observations (\markcite{deBruyn} de~Bruyn 1988) was 5\% per hour at 5 GHz.

Figure~3 displays the
modulation index of PKS~0405--385 as a function of frequency. 
The VLA data for 22~GHz are also plotted. 
The VLA flux densities at
8 and 22~GHz appeared well correlated but the fractional
variability was almost a factor of three smaller at 22~GHz. However,
the interval covered by the VLA was too short to
show the full extent of the variations. We used these data
only to determine the modulation index at 22~GHz as a ratio of modulation indices at 8 and 22~GHz from VLA data, combined with the 8~GHz modulation index obtained from the ATCA data.

Continued monitoring of the radio flux density of PKS~0405--385 at the ATCA
 and HARTRAO
showed that the rapid variability persisted for several weeks
beyond the initial discovery. Remarkably, the variations then
disappeared; they were present at full strength on July 2 
and entirely absent on July 21 . In its steady condition, the
flux density of the source settled on a value which, at each frequency,
was close to the maximum values previously seen during
the variable phase. Analysis of sparse ATCA archival
data indicates that the rapid variability is episodic.

We are currently monitoring PKS~0405--385 on a daily basis.
Our preliminary analysis indicates that the
flux density of the source is steadily rising --- over 10\%
in 2 months starting from 1996 July when the rapid variability ceased. 
The brightening is strongest at 4.8~GHz. 

We were unable to investigate the possible correlation
between the optical and radio variability, as has
been claimed for other IDV sources (\markcite{Quirrenbach} Quirrenbach et al. 1989) due to
inclement weather. No variations
in the optical R-band flux could be seen in data taken with the 2.3 m
telescope of the Mount Stromlo and Siding Spring Observatories in 1996 August,
during a period of radio quiescence. However,
PKS~0405--385 has been reported to vary at mm wavelengths as measured
with the 15 m Swedish ESO Submillimetre Telescope (\markcite{Wagner} Wagner \& Witzel 1995).

\section{Interpretation}
Previous examples of IDV (cf. \markcite{Wagner} Wagner \& Witzel 1995)
stimulated  theories based on either intrinsic mechanisms
or line-of-sight effects. In the latter
category one has gravitational microlensing (\markcite{Wagner} Wagner \& Witzel 1995)
and interstellar scintillation (\markcite{Rickett90}Rickett 1990; \markcite{Rickett95}Rickett et al. 1995), while intrinsically high brightness
temperatures might arise from relativistic shock
waves propagating in a synchrotron source (\markcite{Qian} Qian et al. 1991),
or from coherent (non-synchrotron) radiation processes
(e.g. \markcite{Baker} Baker et al. 1988). 

\subsection{Intrinsic Variability}
To explain an inferred brightness temperature of
${\rm T_B\sim10^{21}\;K}$ in terms of a beamed synchrotron source, 
bulk motion with Lorentz factor $\gamma >10^3$ is required. This is
much larger than any of the values ($\gamma <10$) measured
for superluminal sources with VLBI imaging (cf. \markcite{Vermeulen} Vermeulen \& Cohen 1994).
The difficulties associated with explaining IDV in terms of ultra-relativistic
beaming are considered in detail by \markcite{Begelman94} Begelman, Rees and Sikora (1994). 

Although several
proposals have been made for coherent emission mechanisms (\markcite{Colgate} Colgate 1967;
Baker et al. 1988; \markcite{Lesch} Lesch \& Pohl 1992) there are no entirely satisfactory theories for
such emission in active galaxies and quasars.
In the current picture of how quasars generate their power
-- that is, accretion of matter onto a massive black hole -- any
hypothesised radiation of very high brightness temperature
(${\rm T_B\gg10^{12}\;K}$) is expected to suffer severe attenuation
as a consequence of stimulated electron scattering within
the source (\markcite{Coppi} Coppi, Blandford, \& Rees 1993). 
\subsection{Gravitational Microlensing}
When the line-of-sight to the source passes through a foreground
galaxy, gravitational microlensing can occur as a consequence of
the small-scale structure in the gravitational potential --- due,
for example, to individual stars. For PKS~0405--385, this possibility
is made particularly interesting by the observation of an absorption-line
system, at lower redshift, in the optical spectrum (\S2). However,
our VLBA data show that PKS~0405--385 is not multiply imaged by any
intervening galaxy, so we expect the optical depth to gravitational
microlensing to be small. This means that any microlensing events
should show up as isolated events on an otherwise steady baseline;
moreover, the event time-scale ought to be much longer than is observed
for PKS~0405--385, unless the source moves with an apparent superluminal
transverse speed $>10^2c$.

\subsection{Scintillation}
A second possible extrinsic effect is due to the turbulent motion of the random, small-scale structure in the ionized
interstellar medium (ISM).
Adopting the `standard' theoretical description -- involving a
Kolmogorov spectrum of inhomogeneities, with no inner or outer
scale within the regime of interest -- one arrives at predictions for
the broad-band modulation indices of a point source. These are
(\markcite{Narayan} Narayan 1992): $m_p=(\nu/\nu_0)^{17/30}$ for $\nu<\nu_0$, i.e. in
the `strong' scattering regime, and $m_p=(\nu_0/\nu)^{17/12}$
for $\nu>\nu_0$ (the `weak' scattering regime). (Note, however,
that these are asymptotic results which are not strictly valid in
the regime $\nu\simeq\nu_0$.) Further, from existing models of the
interstellar medium (see particularly \markcite{Taylor} Taylor \& Cordes 1993) we can
estimate the transition frequency for this line of sight to be
$\nu_0\sim5$~GHz. The predicted
modulation index incorporates only one free parameter, the
fraction of the total source flux density in the most compact component ($S_{c}$).
We obtained the best fit of the observed modulation index behaviour assuming $S_{c}= 0.15$ for all frequencies. The results are plotted in Figure 3.
It is clear from this that there is generally good agreement with the data, despite the simple nature of our model of a scintillating point source plus
non-scintillating extended component.

We estimate the angular size of the source, $\theta$, as
follows. The ISM along this line of sight is distributed over a distance
of order 1~Kpc from us. In the weak scattering regime, which is relevant
to our 5 and 8~GHz observations, the source size should be
comparable to, or smaller than, the first Fresnel zone of the equivalent
``phase screen'', in order that substantial variability be observed, whence
$\theta\sim\sqrt{c/(2\pi D\nu)}$. For $\nu = 5$~GHz and adopting a
screen distance of 500 pc, we find $\theta\sim$ 5 $\mu$arcsec. 
This limit is
much smaller than the milli-arcsec dimensions usually found for quasar
cores, and would imply that PKS~0405-385 is the most compact radio quasar
yet found.

In turn, the size of the first Fresnel zone and the observed time scale
for the variations, $\simeq2$~hours, tells us the velocity 
of the inhomogeneities in the ISM relative to the line of sight. 
Here $v\sim\theta D/t\sim50$~km s$^{-1}$ compared with the 30~km s$^{-1}$ of 
the Earth's orbital speed around the Sun.

At frequencies less than 5~GHz the variability time scale increases rapidly 
with wavelength and the modulation index
gradually decreases, as expected for scintillation in the
strong scattering regime. However, the sampling window for our data does not permit us to make good quantitative estimates of the variability timescales at 1.4 and 2.4~GHz.

\section{Discussion and Conclusions}

The observed time scale for
flux density variations is determined by a characteristic transverse
velocity at the distance of the origin of the variations. In the present
model, interstellar scintillations reproduce the observed variability
time scale as a direct consequence of the screen being nearby. Speeds of the order of $50$~km s$^{-1}$ suffice in the ISM where D $\sim500$~pc,
whereas ultra-relativistic motions are implied for variations 
introduced at cosmological distances.

Within the scintillation interpretation we have estimated an upper limit on
the component size of $<5$~$\mu$arcsec, if it has a flux density of 0.24~Jy at 4.8~GHz. This requirement translates to a brightness temperature of
$>5\times10^{14}$~K. In the presence of relativistic motion, the former measurement transforms
as ${\rm T_B^\prime={\cal D}T_B}$ (where ${\cal D}\equiv\nu^\prime/\nu$). Hence
bulk motion with Doppler factor of the order of $10^3$ would be implied for both intrinsic
and scintillation explanations for the variability, if we interpret the
high brightness temperature as being a consequence of relativistic beaming.

The transient behaviour of PKS~0405-385 variability opens up the possibility that the inverse Compton limit might,
for a short time, be greatly exceeded in this source; we note that the
light travel time across a 5 $\mu $arcsec at z$\sim $1 is of the order of a month. In this
circumstance it might not be necessary to have recourse to extreme
levels of relativistic beaming.
We suggest that the transient behaviour is due to the (dis)appearance of
the very compact component in PKS~0405-385.  Assuming this component is a
core with a size of
$<5$~$\mu$arcsec, it will clearly scintillate. The core structure is likely to
evolve, increasing in size to the point where it becomes sufficiently
large that scintillations cease. This could occur
through the ejection of an expanding jet, in which case the flux density
would also likely increase at the same time, thus providing a natural
explanation for the high values of the quiescent flux density observed when
the scintillations ceased in 1996 July.
When the jet becomes transparent at GHz frequencies, the point-like core may appear again, causing the reappearance of
 scintillations. Confirmation of these suggestions will have
to wait until the scintillations reappear and we can follow their evolution
in detail.

Our observations of PKS 0405-385 raise the immediate
question why such extreme variability has been observed in only this one source
out of the large number of known radio quasars? The simple answer may be
that radio quasars are generally not this compact, and
are limited to angular sizes greater than a few tens of $\mu$arcsec at GHz frequencies.
Such variability has
been found only as a result of intensive monitoring campaigns, and could
easily have been missed for other AGNs through less
frequent sampling and through the transient nature of the variability. A definitive
answer to the question will need more extensive monitoring of many more sources.

In summary, we have observed profound variability in the radio quasar PKS
0405-385 on timescales much shorter than anything previously seen in an
extragalactic source. These variations can be understood as due to
interstellar scintillation which, in turn, requires that the source
be very intense, for periods of about a month, with brightness temperature
${\rm T_B > 5\times10^{14}}$~K well above the inverse Compton limit.
Alternative explanations require sources which are sufficiently
small in angular dimensions that they will scintillate in any case (\markcite {Rickett95} Rickett et al.1995). It is evident that PKS~0405-385 will
be a most interesting target for the current Space VLBI missions VSOP
(\markcite{Hirosawa} Hirosawa 1991) and RadioAstron (\markcite{Kardashev} Kardashev \& Slysh 1988).

\acknowledgements

We thank the Director and staff of the VLA for promptly
securing `target of opportunity' observations of this source. We are
also grateful to Don Melrose and Alan Marscher for useful discussions,
Dick Hunstead for his optical line identifications, and
Katrina Sealey for acquiring a confirmation spectrum.

\clearpage

\figcaption[our1.ps]{Three 12 hours series of flux density measurements of PKS~0405--385 between 1996 June 8 and 10 at four frequencies: 8.6, 4.8, 2.4
and 1.4~GHz, from top to bottom, respectively.
 \label{fig1}}

\figcaption[apjpap1.ps]{Combined ATCA and HARTRAO lightcurve measured on the 1996 June 8 at (a) 8.6/8.4~GHz (ATCA/HARTRAO) and (b) 4.8/5.0~GHz (ATCA/HARTRAO); ATCA data are plotted as circles ($\sigma = 0.01$ Jy), HARTRAO data as triangles ($\sigma = 0.05$ Jy).
 \label{fig2}}

\figcaption[fig4.ps]{The modulation index -- i.e. the fractional rms variations
-- as a function of frequency. Also shown is a one-parameter theoretical
fit for the expected scintillation behaviour (the free parameter is the
fraction of the source flux density in the compact component). \label{fig3}}
 
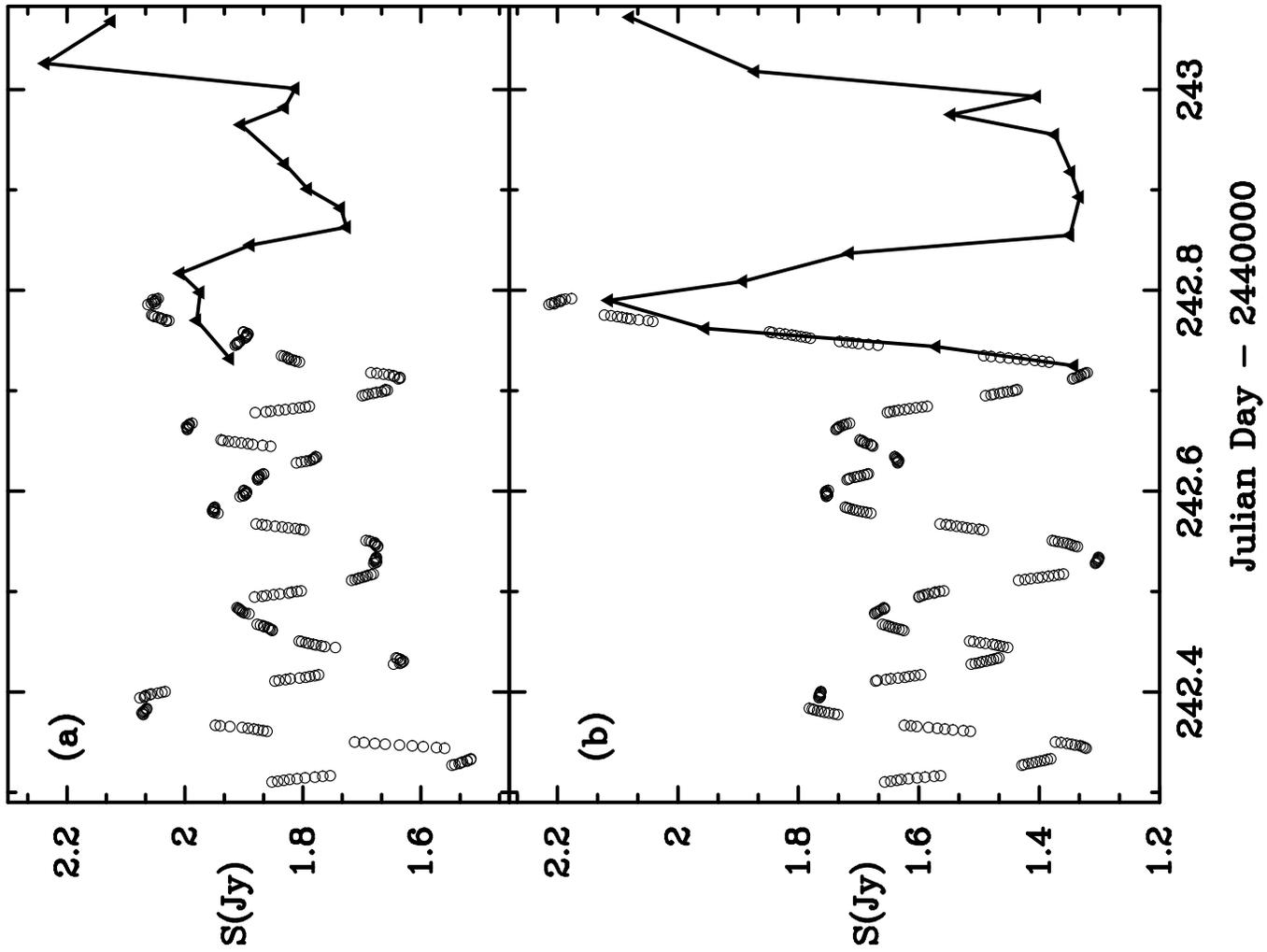
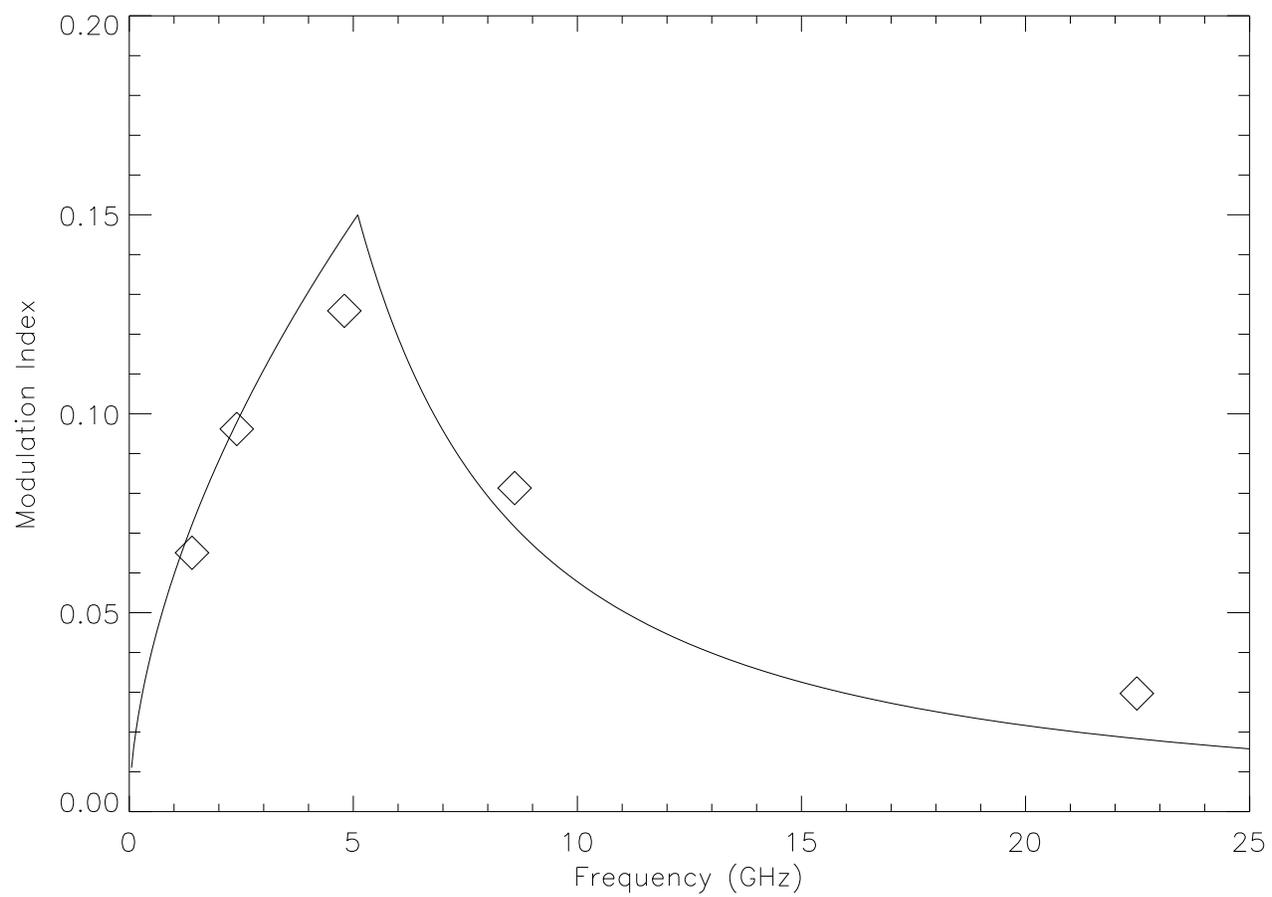
\end{document}